\documentclass[11pt]{article}

\usepackage{natbib}
\usepackage{graphicx}
\usepackage{hyperref}
\usepackage[hyphenbreaks]{breakurl}

\renewcommand{\section}{\subsection}

\begin{document}

\centerline{\LARGE\bf Crowdfunding Astronomy Research with Google Sky}

\vspace*{11pt}
\centerline{Travis S.\ Metcalfe, White Dwarf Research Corporation, USA}
\vspace*{22pt}

\centerline{\bf ABSTRACT}
\vspace*{11pt}

\noindent{\it For nearly four years, NASA's Kepler space telescope 
searched for planets like the Earth around more than 150,000 stars similar 
to the Sun. In 2008 with in-kind support from several technology 
companies, our non-profit organization established the {\sl Pale Blue Dot 
Project}, an adopt-a-star program that supports scientific research on the 
stars observed by the Kepler mission. To help other astronomy educators 
conduct successful fundraising efforts, I describe how this innovative 
crowdfunding program successfully engaged the public over the past seven 
years to help support an international team in an era of economic 
austerity.}

\vspace*{11pt} 
\noindent{\bf Keywords:} Astronomy Education; Kepler; Pale Blue Dot 
Project; Adopt a Star; Crowdfunding

\vspace*{22pt}
\noindent The Kepler space telescope was launched in 2009 and spent the 
next four years monitoring the brightness of more than 150,000 stars in 
the constellations Cygnus and Lyra. The mission was designed not only to 
discover planets around other stars by searching for periodic dips in 
brightness caused by fortuitously aligned orbits, but also to characterize 
the parent stars of the newly discovered planetary systems using stellar 
seismology. As a cost-saving measure prior to launch, in 2004 NASA dropped 
all financial support for stellar seismology and forged a partnership with 
a large collaboration led by scientists in Denmark \citep{Gilliland2015}. 
In exchange for influence on the selection of targets and early access to 
proprietary observations, the team provided data analysis services at no 
cost to NASA. The high-profile agreement gave many European scientists an 
advantage to secure funding from their home countries, but left US 
participants largely without support.

As an early-career researcher in 2007, I submitted a proposal to NASA's 
Kepler Participating Scientist program to help fund my involvement in the 
Danish-led collaboration. The idea behind the program was to augment the 
Kepler science team with new expertise that could significantly enhance 
the scientific return of the mission. Although a panel of independent 
reviewers recommended my proposal for funding, the Kepler science team 
decided not to support it. Seeking an alternative source of support for 
our work, in January 2008 we started the {\sl Pale Blue Dot 
Project} \citep[\url{http://adoptastar.whitedwarf.org};][]{Metcalfe2009}.

\begin{figure}[t]
\centerline{\includegraphics[angle=0,height=4.0in]{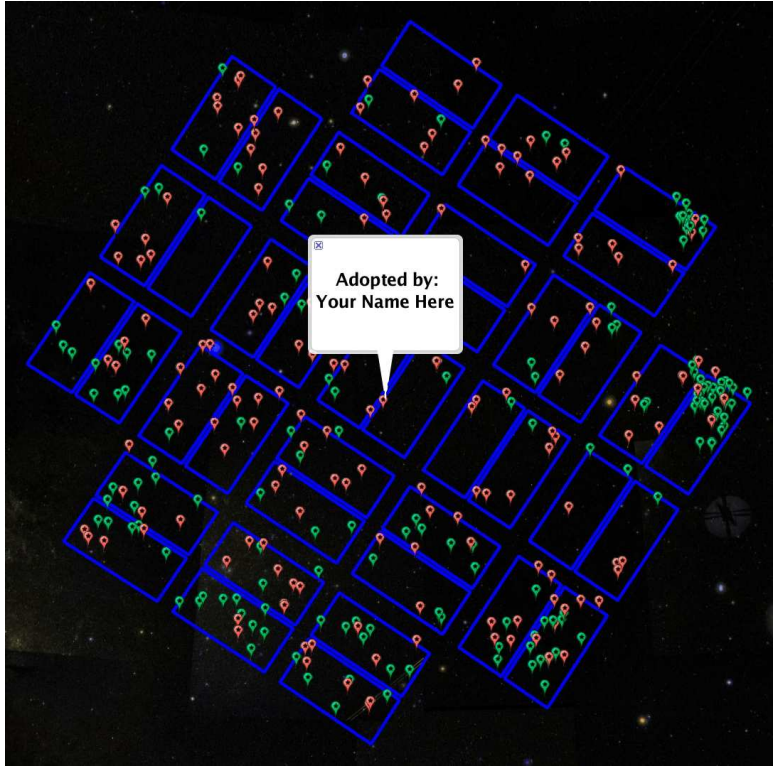}}
\caption{A representation of the Kepler target list in Google Sky. 
Fainter targets are marked as the user zooms in to individual regions of 
the telescope field of view (blue boxes). Adopted stars are marked with 
red icons, while available targets are marked with green icons.\label{fig1}} 
\end{figure} 

\vspace*{11pt}
\centerline{\bf GUIDING PRINCIPLES \& IMPLEMENTATION}
\vspace*{11pt}

For-profit companies have been offering informal ``name a star'' services 
for decades, the most notable being the International Star Registry. In 
stark contrast, the {\sl Pale Blue Dot Project} was designed to compete 
against these companies while adhering to several guiding principles more 
consistent with tacit perspectives of the broader scientific community. 
First, we wanted to be intellectually honest about what we were providing, 
so we branded our program an ``adopt a star'' service to dissuade the 
notion that we were somehow selling stars. This approach is analogous to 
an ``adopt a highway'' program that seeks local sponsors to maintain clean 
roads, where it is well understood that the sponsor does not actually own 
the road or get to rename it. Instead, a sign along the side of the road 
acknowledges their support. Second, our program is run by a non-profit 
organization with 100\% of the proceeds supporting astronomy research. 
This was only possible with in-kind support from several technology 
companies and thousands of volunteer hours over the past seven years. 
Finally, the program sought to establish a direct link between the 
fundraising mechanism and the science that it supports. Only the Kepler 
target stars are available for adoption, and the funds generated by the 
program support research on that same set of stars.

Now in its eighth year of operations, the {\sl Pale Blue Dot Project} 
allows anyone to adopt a Kepler target star for a \$10 USD donation to 
support scientific research. Each donor receives a personalized 
``Certificate of Adoption'' by email, with a name or dedication of their 
choice and a unique star number from the Kepler Input Catalog 
\citep{Brown2011}. We update the target on our website with the name of 
the donor, ensuring that each star can only be adopted once. The Kepler 
target list was not publicly available when we started the program, so we 
encouraged ``early adopters'' to donate without selecting a specific star 
and gave them the first choice of targets as soon as the catalog was 
released.

The centerpiece of our outreach and crowdfunding effort was a 
representation of the Kepler target list in Google Sky, allowing potential 
donors to browse the locations and basic properties of the stars that NASA 
was searching for planets. As shown in Figure~\ref{fig1}, Google Sky uses 
a markup language to project icons and other information on top of 
high-resolution images derived mostly from digitized Palomar Sky Survey 
photographic plates \citep{Connolly2008}. Any attempt to display all 
150,000 targets simultaneously would slow down the computer and ultimately 
crash the browser. What we needed was a recipe for initially marking the 
brightest stars, and then gradually highlighting the fainter stars as the 
user zooms into the field. The open-source regionator software 
(\url{https://code.google.com/p/regionator/}) developed at Google provided 
exactly the functionality that we needed. It was even distributed with a 
sample application to display the Swiss national rail network in Google 
Earth, starting with connections between the busiest train stations but 
eventually showing all of the connections as the user zooms in. We simply 
replaced train stations with stars, and used brightness to prioritize the 
targets.

\vspace*{11pt}
\centerline{\bf SUCCESSES \& CHALLENGES}
\vspace*{11pt}

\begin{figure}[t]
\centerline{\includegraphics[angle=0,height=4.0in]{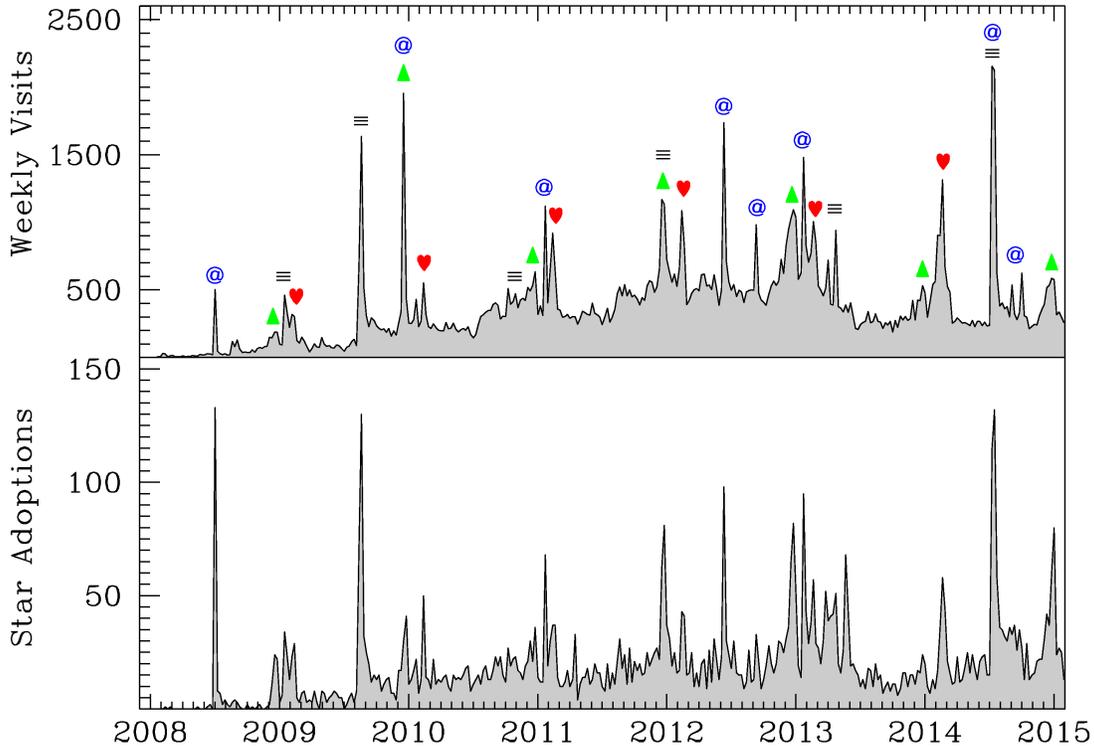}}
\caption{Weekly visits and star adoptions during the first seven years of 
our crowdfunding campaign. Icons mark significant peaks in web traffic 
from traditional media coverage ($\equiv$), social media exposure (@), and 
annually during Christmas ($\Delta$) and Valentine's Day ($\heartsuit$).
\label{fig2}}
\end{figure}

Over the past seven years the {\sl Pale Blue Dot Project} has gradually 
grown and is attracting hundreds of donations each month. The history of 
weekly website visits and star adoptions is shown in Figure~\ref{fig2}. 
For the purposes of the figure, a visit is defined as a session with a 
duration of up to 30 minutes and at least one interaction (e.g.\ clicking 
a link). Sessions that access a single page without interaction typically 
outnumber these visits by a factor of four. Star adoptions account for the 
total number of successful donations without regard to their value. In 
2012 we began offering star adoptions with different donation amounts for 
value-added targets: \$15 for double stars from the Kepler eclipsing 
binary catalog \citep{Prsa2011}, \$25 for stars with suspected planets 
from the NASA exoplanet archive 
(\url{http://exoplanetarchive.ipac.caltech.edu/}), and \$100 for stars 
with confirmed planets and an official Kepler number. Prior to 2012, our 
strategy involved driving more traffic to the website to attract more 
donations. Since then our focus has shifted to converting a higher 
fraction of our visitors into donors.

The history of our crowdfunding program is punctuated by recurring 
features and episodic spikes in both visits and star adoptions. The 
recurring features correspond to popular gift-giving holidays (Christmas 
and Valentine's Day). The episodic spikes arise from two sources of 
publicity: (1) traditional media coverage of our program, generally 
associated with published scientific results that we supported, and (2) 
social media exposure, often initiated by enthusiastic donors. Although 
there is a strong correlation between visits and donations, the dependence 
is clearly non-linear. Strategically timed publicity can raise awareness 
of the program during the already-popular seasons. Sometimes this leads to 
an increase in visits without a comparable boost in donations (as with the 
social media exposure in late 2009), while other times it has a 
disproportionate impact on donations (as with the traditional media 
coverage of \cite{Fellet2011}).

There is an interesting story behind almost every spike in star adoptions, 
and it is worth elaborating on the largest few because they were 
associated with some of the greatest challenges to our program. The first 
spike in mid-2008 was the result of a short post on the tech news 
\nobreak{website Slashdot} 
(\url{http://science.slashdot.org/story/08/07/01/1724256/adopt-a-star-to-fund-research}). 
In 2008 the concept of crowdfunding was still in its infancy, and popular 
platforms like Kickstarter didn't yet exist. This novelty and the strong 
appeal to the interests of Slashdot readers led to the highest rate of 
donations in the history of the project, with roughly one in three 
visitors proceeding to adopt a star. We had certainly found our 
constituency, but we also attracted the attention of the Kepler science 
team.

Somewhat unexpectedly, a few days after the Slashdot post appeared, we 
received an email from NASA's Principle Investigator for the Kepler 
mission, William Borucki. He expressed some concerns about whether we 
planned to allow star adoptions only for the Kepler targets, and 
questioned our intentions for how the funding would be used. He requested 
that we include a disclaimer on the website to inform visitors that our 
program was ``not authorized or endorsed by NASA''. We added a disclaimer 
to the project website, but the controversy prompted one member of our 
board of directors to resign, and we ultimately had to reconstitute the 
board with younger colleagues who were more comfortable with the concept 
of crowdfunding.

The second large spike in star adoptions came from traditional media 
coverage of a press release about our program in August 2009. An initial 
article was published two days after the press release by Space.com 
\citep{Moskowitz2009}, with an additional article by New Scientist 
\citep{Grossman2009} a few days later. Both articles included links to our 
website, making it easy for readers to support the program. The impact of 
this media coverage exceeded the Slashdot post because the response was 
sustained over a longer period of time. Shortly after the New Scientist 
article appeared, we received an email from a representative of the Carl 
Sagan Foundation. The article had included the following sentence: {\sl 
``The programme ... is called `Pale Blue Dot' to echo Carl Sagan's 
description of Earth as seen from space.''} The representative insisted 
that the name of our project constituted an unauthorized use of 
copyrighted material, since `Pale Blue Dot' was the title of a book 
written by Carl Sagan in 1994. Six weeks later, we received a ``cease and 
desist'' letter from an attorney representing the Sagan estate. We spent a 
month working with a {\sl pro bono} lawyer to refute their illegitimate 
claim to the phrase (book titles are not protected by copyright law).

\begin{figure}[t]
\centerline{\includegraphics[angle=0,height=4.0in]{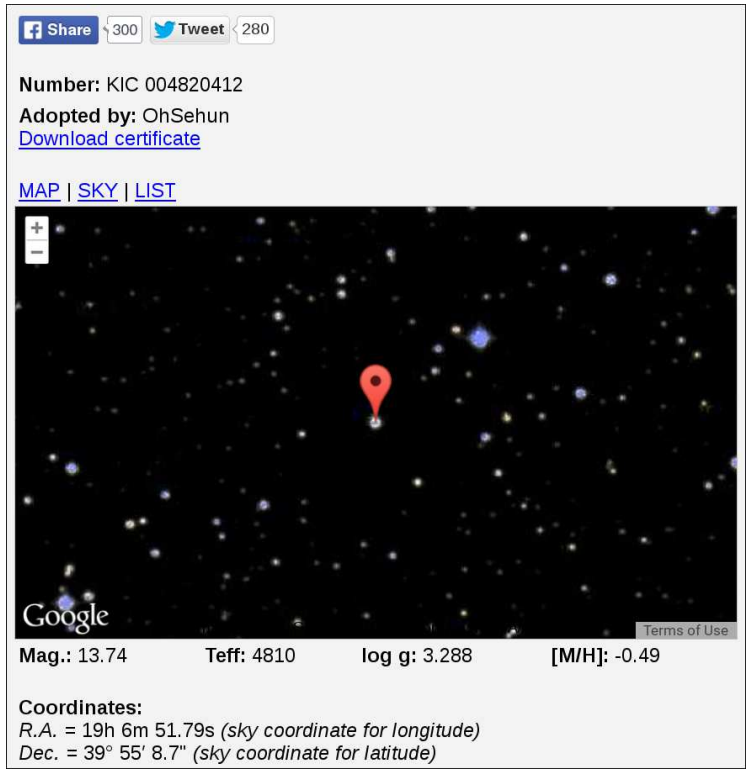}}
\caption{An example star page from the {\sl Pale Blue Dot Project} database, 
where users can share the star adoption on social media, download their 
``Certificate of Adoption'', view an image of the star in Google Maps or 
Sky, and see a list of nearby stars to adopt for friends.\label{fig3}}
\end{figure}

The largest surge in star adoptions during the history of our project came 
in the middle of 2014, when Ukrainian astronomers dedicated a star to an 
unflattering nickname for Russian president Vladimir Putin. As intended, 
our \$10 USD star adoptions were undercutting the prices of for-profit 
competitors. This motivated independent operators in several countries to 
market Kepler star adoptions in their native language, adopting stars 
through our website and then reselling them to their customers with 
additional services. A political group in Ukraine had adopted a star under 
the controversial name through one of these operators, and then began 
publicizing it through social media in late June 2014. By early July the 
insult had gone viral, with an image of the Ukrainian ``Certificate of 
Adoption'' guiding supporters to the {\sl Pale Blue Dot Project}. On the 
morning of July 4 that year, I received a telephone call from a reporter 
at the Moscow Times 
(\url{http://www.themoscowtimes.com/news/article/no-plans-to-rename-star-called-putin-is-a-dckhead/503017.html}) 
asking for comment. I told him {\sl ``Free speech is now written in the 
stars. We have no plans to censor any of these star adoptions. We 
appreciate the support for science.''} Several patriotic Russians were 
disappointed with our decision.

\vspace*{11pt}
\centerline{\bf LESSONS LEARNED}
\vspace*{11pt}

Although the Google Sky interface to the Kepler target list serves as a 
highly useful public outreach tool, the interface sometimes proved to be a 
frustrating technical obstacle for many visitors. To view it through our 
website, users were required to install the Google Earth plugin on their 
Internet browser. The plugin was available for the most common desktop 
operating systems, but there was no support under Linux or for mobile 
devices. Looking for an easier way to select a star, Dutch software 
developer Sjors Provoost imported the Kepler target list into a database 
and created a simple interface to search it and to sort the results. One 
consequence of his approach was that each star in the database had its own 
webpage, where he included an image of the star using Google Maps along 
with coordinates and other basic properties from the Kepler Input Catalog. 
When he finished, he uploaded the software to an open-source repository 
(\url{https://github.com/Sjors/pale-blue-dot/tree/master}) and helped us 
reconfigure it for our program. We added social media share buttons, a 
link to the ``Certificate of Adoption'', the option to switch between 
Google Maps and Sky, and a list of nearby Kepler targets. An example is 
shown in Figure~\ref{fig3}, the most-shared star in our database, which 
was adopted for a member of the Korean pop band EXO.

The impact of offering an easier method of selecting a star through the 
database reinforced the most important lesson we have learned from our 
crowdfunding program: keep it simple. Most of our visitors seem to fall 
into one of two main categories: (1) ``enthusiasts'' who are excited about 
the science we do and are happy to support us, and (2) ``customers'' who 
are looking for a unique gift delivered immediately at a low price. The 
first group is interested enough to read the text on our website, browse 
the Kepler stars in Google Sky, and support our science with larger 
donations for the value-added targets. The second group wants the process 
to be intuitive enough to complete by following visual cues, they are not 
so interested in the specific star they adopt, and they are less likely to 
donate more than the minimum. Serving the needs of both groups has made 
our program more appealing to all visitors. We minimize text up front, but 
we include links to more information for those who want additional 
details. Access to the database makes it easy to select a star with 
specific characteristics, but we assign the brightest star available by 
default so no selection is required. Value-added targets create an 
incentive for generosity, but large numbers of small donations still help.

A generation ago, it was not uncommon for scientists to rely on other 
professionals for some of the things that we now do for ourselves. For 
example, in the past, there were software engineers to write computer 
code, drafters to create camera-ready figures, secretaries to edit and 
publish manuscripts. Today, desktop computers make it easier for 
scientists to be more self-reliant, but the fact that we can learn how to 
do web development, database management, and online marketing doesn't make 
us the most qualified for the job. The final lesson we learned from the 
{\sl Pale Blue Dot Project} is to hire consultants for the most important 
work. We brought in an experienced web developer to redesign the website 
in 2014, and annual donations nearly doubled. We are now working with a 
professional marketing firm to overhaul our online advertising strategy. 
In the first seven years of the program we attracted about \$100,000 in 
donations, but we brought in half of this total in just the past two 
years. Relatively small investments in people who do not think like 
scientists can reap benefits that quickly exceed their cost, giving us 
more time and funding for our research.

\vspace*{11pt}
\centerline{\bf FUTURE OUTLOOK}
\vspace*{11pt}

When we started the {\sl Pale Blue Dot Project} seven years ago, we 
envisioned quickly raising \$1.5 million USD and establishing an endowment 
for our research over the lifetime of the Kepler mission. In a different 
universe, it might have played out that way. One of our ``early adopters'' 
in 2008 knew a writer for the television show {\sl The Colbert Report}, 
and pitched an idea for a segment about our project. In 2010 we were 
contacted by a marketing firm looking to celebrate the 25$^{\rm th}$ 
anniversary of the {\sl Super Mario Brothers} video game by adopting a 
constellation of Kepler stars in the shape of Mario's face. Neither of 
these opportunities came to fruition, but we still raised enough funding 
over the years to help students and early-career scientists present their 
research at annual workshops, and to help some of our colleagues in 
developing countries pay the publication charges for their research 
papers.

With all of the infrastructure in place and a well-established brand, 
there is enormous potential for future missions to benefit from our 
crowdfunding program. The Transiting Exoplanet Survey Satellite (TESS) 
will launch in 2017, and it will do for the brightest stars in the sky 
what Kepler did for one small patch of the summer Milky Way. The European 
PLATO mission will do even more for stellar seismology starting in 2024, 
with extended observations all around the Galactic plane. Both missions 
will rely on international teams to help analyze the observations. With 
foresight and cooperation from these missions, citizens around the globe 
will be able to support this exciting work through the {\sl Pale Blue Dot 
Project}.

\centerline{\bf ACKNOWLEDGMENTS} 
\vspace*{11pt}

\noindent The {\sl Pale Blue Dot Project} has received in-kind support 
from Google, PayPal, Dreamhost, and SmartClick Adworks. We would like to 
thank Sjors Provoost, Ian Shorrock, and Robert Piller for contributions 
that substantially improved our crowdfunding program. This article would 
not have been possible without support from the Stellar Astrophysics 
Centre at Aarhus University in Denmark.

\vspace*{11pt}
\centerline{\bf AUTHOR INFORMATION} 
\vspace*{11pt}

\noindent {\bf Travis S.\ Metcalfe} is the director of White Dwarf 
Research Corporation, and a research scientist based in Boulder, Colorado. 
He started the {\sl Pale Blue Dot Project} in 2008, and he has volunteered 
thousands of hours for daily operations over the past seven years. He 
hopes to see crowdfunding support for more astronomy projects in the 
future, and he is open to collaboration with like-minded individuals.

\end{document}